# Weak localization and anti-localization in rare earth doped topological insulators

Zengji Yue[1,4]*, Kirrily Rule[2,4], Zhi Li[1,4], Weiyao Zhao[1,4], Lina Sang[1,4], Guangsai Yang[1], Cheng Tan[3,4], Lan Wang[3,4], Abuduliken Bake[1], and Xiaolin Wang[1,4]*

1. Institute for Superconducting & Electronic Materials, Australian Institute of Innovative Materials, University of Wollongong, Wollongong, NSW 2500, Australia

2. Australian Nuclear Science and Technology Organisation, Lucas Heights, NSW, 2234 Australia

3. School of Science, RMIT University, Melbourne, VIC 3000, Australia

4. ARC Centre for Future Low-Energy Electronics Technologies (FLEET), University of Wollongong, Wollongong, NSW 2500, Australia

Email: Zengji@uow.edu.au; xiaolin@uow.edu.au;

We study magneto-transport phenomena in two rare-earth doped topological insulators, $Sm_xFe_xSb_{2-2x}Te_3$ and $Sm_xBi_{2-x}Te_2Se$ single crystals. The magneto-transport behaviours in both compounds exhibit a systematic crossover between weak anti-localization (positive magnetoresistance) and weak localization (negative magnetoresistance) with changes in temperatures and magnetic fields. The weak localization is caused by rare-earth-doping induced magnetization, and the weak anti-localization originates from topologically protected surface states. The transition between weak localization and weak anti-localization demonstrates a gap opening at the Dirac point of surface states in the quantum diffusive regime. This work demonstrates an effective way to manipulate the magneto-transport properties of the topological insulators by rare-earth element doping. Magnetometry measurements indicate that the Sm-dopant alone is paramagnetic, whereas the co-doped Fe-Sm state has short-range antiferromagnetic order. Our results hold potential for the realization of exotic topological effects in gapped topological insulator surface states.

Topological insulators (TIs) are one type of topological matter, which have an insulating bulk state and a topologically protected metallic surface state with spin and momentum helical locking and a Dirac-like band structure.[1,2] Backscattering in the Dirac fermions by nonmagnetic impurities is prohibited due to their protection by time-reversal symmetry (TRS). Unique and fascinating electronic properties, such as the quantum spin Hall effect (QSHE), topological magnetoelectric effect, giant magnetoresistance, and Majorana fermions, are expected from topological insulator materials.[3,4] Topological insulator materials also exhibit a number of excellent optical properties, including ultrahigh bulk refractive index, near-infrared frequency transparency, unusual electromagnetic scattering, and ultra-broadband surface plasmon resonances.[5-8] These excellent electronic and optical properties make topological insulator materials suitable for designing various advanced electronic and optoelectronic devices.[9-11] To date, a series of well-known thermoelectric semiconductors, including $Bi_2Se_3$, $Bi_2Te_3$, and $Sb_2Te_3$, have been identified as typical topological insulator materials.[12] Very recently, three groups carried out comprehensive first-principle calculations and found 3,307 TI materials and 8,056 topological materials.[13,14] Furthermore, they revealed that more than 27 per cent of materials in nature are topological.[15]

Magnetic doping can open an energy gap at the Dirac point in TIs due to the breaking of time reversal symmetry by magnetic impurities.[16] A variety of exotic topological effects including the quantum anomalous Hall effect (QAHE), topological magneto-electric effects, image magnetic monopoles, and Majorana fermions can be created through magnetic doping. The QAHE has been predicted and observed in a magnetic topological insulator [17,18]. The spontaneous magnetic moments and spin-orbit coupling in magnetic topological insulators lead to the quantum Hall effect (QHE) without an external magnetic field. The QAHE has demonstrated potential applications in novel electronics with low power consumption. At present, however, the QAHE in magnetic topological insulators is limited to extremely low temperatures. A high-temperature QAHE would be extremely attractive for device applications, however, further work is needed to optimise the magnetic and electronic doping in this family of materials.

Magneto-transport measurements represent an important approach to explore the electronic characteristics of quantum materials.[19-25] Magnetoresistance (MR = $(R_H-R_0)/R_0 \times 100\%$) is the change of electrical resistance of a material in an externally-applied magnetic field. has wide applications in magnetic data storage, magnetic sensors, and magnetoelectronic devices.[26-28] Chalcogenide insulators (eg. $Sb_2Te_3$, $Bi_2Te_3$ ) have large linear giant magnetoresistance over several hundred percent.[3] In magnetically-doped topological insulators, a particularly interesting phenomenon is the cross-over occurs between negative to positive MR with temperature, attributed to the trade-off between weak localization and weak-anti localization effects [29]. To date, however, this has only been reported in transition-metal doped $Bi_2Se_3$ and there is little information about the effect of other dopants, for example, the exotic rare-earth dopants. With the large magnetic moment, rare-earth elements have been widely used in magnets and electronic devices. Unlike defect induced magnetism, rare-earth element doping can generate intrinsic ferromagnetism.[30] The challenge for rare-earth doping is that the exchange interactions between 4f ions tends to be weak, and mediated by electron carriers, generally leading to low Curie/Neel temperatures. A viable strategy to enhance the magnetic properties of rare-earth magnets is, therefore, to combine rare-earth ions with a transition metal component which can foster high Curie temperatures and hard magnetic properties (eg. NdFeB). In this work, we report magnetoresistance transport studies on samarium (Sm) doped TI single crystals and compare therese with Fe-Sm doping. With increasing temperature and magnetic field, the MR exhibits a transition between negative (weak localisation, WL) and positive (weak antilocalisation, WAL) values. The transition can be explained based on the transformation between a TI and a dilute magnetic semiconductor (DMS). The magnetoresistance cross-over may also be a sign of the opening of an energy gap in the surface states. The results demonstrate that rare earth elements are promising dopants for tuning both the electronic band structures and magnetic properties of topological insulators.

High-quality single crystals were grown using the Bridgman method starting with high-purity Sm, Fe, Sb, Te, Bi, and Se elements in the nominal starting concentrations to yield $Sm_xFe_xSb_{2-2x}Te_3$ and $Sm_xBi_{2-x}Te_2Se$ ($x \approx 1\%$). The elements were sealed in a quartz tube, heated to 750 ºC, and kept for 6 hours. Then, the temperature was reduced to 700 ºC and slowly cooled down to 560 ºC at the rate of 2 ºC/h. Finally, the temperature was cooled down to room temperature naturally. The resulting ingots were cracked and cleaved to reveal single crystals. The quality of the bulk crystals was checked using X-ray diffraction (XRD), SEM and EDS. Energy dispersive spectroscopy measurements in SEM reveal that the local concentration of the dopant was slightly inhomogeneous across the ingot. This indicates that excess dopants were expelled from the crystal during the growth process and the solid-solution limit of Sm / Fe is below 0.1 in the $Sb_2Te_3$ matrix giving real doping concentrations << 1 atomic percent. Figure 1a shows the X-ray diffraction (XRD) pattern of $Sm_xFe_xSb_{2-2x}Te_3$ single crystals, which demonstrates the high quality of crystals with growth orientation along the [001] direction. Figure 1b shows the X-ray diffraction (XRD) pattern of $Sm_xBi_{2-x}Te_2Se$ single crystals, which demonstrates the high quality of the crystals.

Electronic transport measurements were conducted in a Physical Properties Measurement System (PPMS, Quantum Design) using the standard four-contact method. The resistances were measured under varying magnetic fields and temperatures.

*Magnetotransport:* $Sb_2Te_3$ is typical topological insulator with *p*-type carriers in the bulk state. Figure 2a shows the temperature dependence of the resistance in doped $Sm_xFe_xSb_{2-2x}Te_3$ single crystals, which present a metallic character. Actually, in the electronic structure of ideal topological insulators, the Fermi level falls within the bulk band gap, which is traversed by topologically-protected surface states. In this case, the material bulk should be insulating. The Fermi surfaces of most real topological insulators are often located in the valence band or conduction band. The intrinsic defects of Sb and Te also contribute a large amount of bulk carriers, which lead to metallic bulk states. In our samples, the measured metallic bulk conductance results from mixed contributions of the bulk states and the surface states. For magneto-transport, Figure 3a shows the MR in $Sm_xFe_xSb_{2-2x}Te_3$ crystals as a function of magnetic field. At 5 K, 10 K and 25 K, the MR remains negative up to 14 T and has a transition in 11 T. With increasing temperature, the MR makes a transition from negative to positive at 50 K and 100 K. The negative MR is caused by the WL and is evident at low temperatures. The WL is related to the magnetization and the degree of time reversal symmetry (TRS) breaking.

We also studied another bulk insulating topological insulator, $Bi_2Te_2Se$ single crystals.[31] The rare earth Sm was doped into this compound and led to $Sm_xBi_{2-x}Te_2Se$. Due to the bulk insulating states in $Sm_xBi_{2-x}Te_2Se$, the surface Fermi level ($E_F$) is located in the bulk band gap, and the transport properties of the gapped surface can be investigated. Figure 2b shows the temperature dependence of the resistance in $Sm_xBi_{2-x}Te_2Se$ single crystals, which presents a metallic character at higher temperature (>100 K) and insulating properties at lower temperature (< 100 K). Similar semimetal-semiconductor transition has been observed in three-dimensional (3D) Dirac semimetals.[32] Figure 3b shows the MR in $Sm_xBi_{2-x}Te_2Se$ single crystals as a function of temperature and magnetic fields. The MR demonstrates a transition from positive to negative with increasing magnetic field. At 2.5 K – 50 K, the MR is negative at low magnetic fields and becomes positive at high fields. This is caused by a transition from WAL to WL. Above 50 K, the MR remains positive and has a parabolic shape.

*Magnetometry:* Figure 4 presents the magnetization of $Sm_xFe_xSb_{2-2x}Te_3$ single crystals as a function of temperatures. The $Sm_xFe_xSb_{2-2x}Te_3$ single crystals show antiferromagnetism (AFM) with a Néel temperature ($T_N$) of 125 K. Both samples show similar magnetic saturation, consistent with a very low doping concentration. For comparison a 3% Fe sample would be expected to yield nominally ~ 1 emu/gram, and the magnetometry approximately implies a ten-fold reduction in the true doping concentration. The magnetization results correspond to the magnetotransport results. When the material is antiferromagnetic, it has a negative MR and WL. When the material is paramagnetic, it shows a positive MR and WAL. The $Sm_xBi_{2-x}Te_2Se$ single crystal displays paramagnetism over all the temperatures. In this case, the spin moments are easily aligned and the magnetization could lead to negative MR.

Rare-earth doping into topological insulators is one way to introduce higher magnetic moments into magnetic topological insulators. Dy, Ho, and Gd have been used as dopants in topological insulator thin films.[33] High magnetic moments can be introduced into the topological insulators, but it is difficult to achieve ferromagnetic order. The high moment rare-earth ions behave like isolated magnetic moments and lead to overall paramagnetic behavior in the doped topological insulators. To date, long-range ferromagnetic order has only been observed in rare-earth-doped magnetic topological insulator heterostructures due to the proximity effects and interface effects.[34] Ferromagnetic ordering with a Curie temperature ($T_C$) up to 17 K has been found in the Dy doped layer in Cr doped $Sb_2Te_3$ and Dy doped $Bi_2Te_3$ thin film heterostructures. This demonstrates that a high moment of the rare earth can have long-range magnetic order through heterostructure engineering. Nevertheless, a band gap can still be opened in the topological surface states in the absence of long-range ferromagnetic order. Short range ferromagnetic order induced by inhomogeneous magnetic doping and the formation of rare-earth clusters is responsible for the gapped topological surface states.[35]

In our work, we observed a transition between WL and WAL in rare-earth-doped topological insulators. The MR demonstrates quite different behavior in $Sm_xFe_xSb_{2-2x}Te_3$ and $Sm_xBi_{2-x}Te_2Se$ single crystals. They show opposite transitions with increasing magnetic field. Observation of the crossover is an important signature for the existence of a TRS-breaking gap in the topological surface states.[36] The WL results from Sm doping-induced magnetization in $Sm_xFe_xSb_{2-2x}Te_3$ and $Sm_xBi_{2-x}Te_2Se$ single crystals. WL is due to the constructive quantum interference between the two loops. The WAL behavior originates from strong spin-orbit-coupling (SOC) in the bulk and spin-momentum locking in the topological surface states [37]. WAL has been observed in the topological insulators $Bi_2Se_3$ and $Bi_2Te_3$, which originates from topologically protected surface states [38,39]. The surface Dirac electrons travel along two time-reversed self-crossing loops and accumulate a Berry phase. The destructive quantum interference between them reduces the return probability of the Dirac fermions, leading to a quantum enhancement of the classical Drude conductivity [39].

Magnetically doped topological insulators can undergo a WL to WAL crossover due to the TRS-breaking gap opened at the Dirac point of the topological surface states[36]. The magnetization-induced WL competes with the SOC induced WAL. The WL to WAL crossover in $Sm_xBi_{2-x}Te_2Se$ is due to the transformation of the system from a topological insulator (with WAL) to a topologically trivial diluted magnetic semiconductor (with WL) driven by magnetic impurities. The crossover from WAL to WL is driven by a certain type of symmetry breaking that suppresses the topological protection. Lu et al. revealed a competing effect of WL and WAL in quantum transport when an energy gap is opened at the Dirac point by magnetic doping.[36] This crossover presents a unique feature characterizing the topological surface states by the gap opened at the Dirac point in the quantum diffusion regime. Antiferromagnetism, unlike ferromagnetism, does not break time reversal symmetry in normal circumstances. Consequently, the low-field MR does not exhibit a change in sign at temperatures, indicating no obvious cross-over from WAL to WL. On the other hand, the application of a large magnetic field can cause antiferromagnetic spins to cant and break time-reversal symmetry. This may be the origin of the inversion in the MR sign at ~5 T below 20 K

In summary, we measured the magneto-transport properties in Sm doped $Sb_2Te_3$ and $Bi_2Te_2Se$, and observed a complex crossover between WAL and WL. The transition demonstrates the suppression of the nontrivial bulk topology by magnetic rare-earth doping. The transition is also evidence of surface gap opening in rare-earth Sm-doped topological insulators.. The results showing the possibility of short-range antiferromagnetic order due to the co-doping effect of Fe-Sm. The tuning of electron transport by rare-earth doping paves the way to finding exotic topological effects.

Acknowledge: We thank Dr. David Cortie for helpful discussion and modification. We thank Dr. Guolin Zheng, A/Prof. Dimitrie Culcer and Prof. Michael Fuhure for valuable discussion.

Figure Captions

Fig. 1. XRD patterns of t (a) $Sm_xFe_xSb_{2-2x}Te_3$ and (b) $Sm_xBi_{2-x}Te_2Se$ single crystals.

Fig. 2. Resistance as a function of temperatures in (a) $Sm_xFe_xSb_{2-2x}Te_3$ and (b) $Sm_xBi_{2-x}Te_2Se$ single crystals.

Fig. 3. MR in (a) $Sm_xFe_xSb_{2-2x}Te_3$ and (b) $Sm_xBi_{2-x}Te_2Se$ single crystals at different temperatures.

Fig. 4. (a) Magnetization of $Sm_xFe_xSb_{2-2x}Te_3$ single crystals as a function of temperature. (b) Magnetization of $Sm_xBi_{2-x}Te_2Se$ single crystals as a function of temperature.

Fig. 5. (a) Magnetization of $Sm_xFe_xSb_{2-2x}Te_3$ single crystals as a function of magnetic fields. (b) Magnetization of $Sm_xBi_{2-x}Te_2Se$ single crystals as a function of magnetic fields.

Figures

Figure. 1

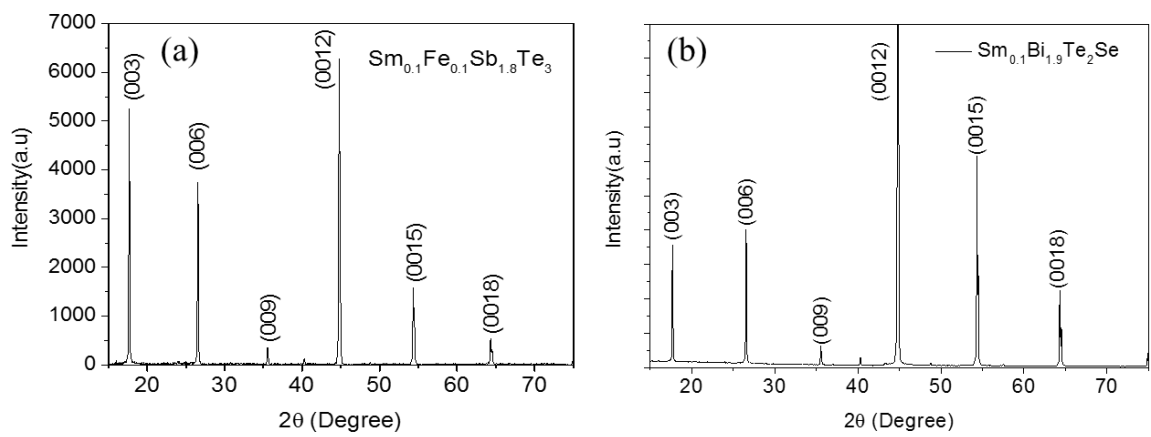

Figure. 2

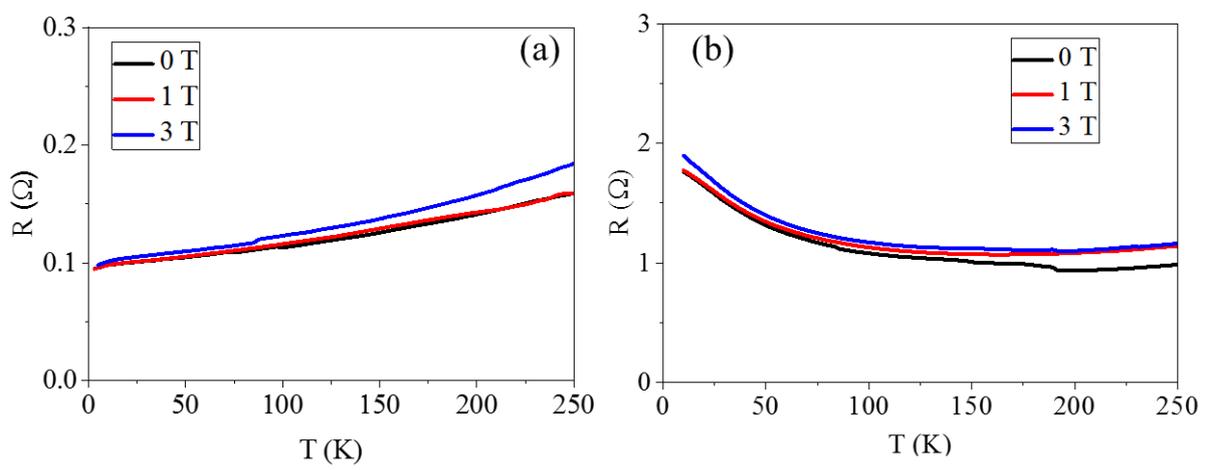

Figure. 3

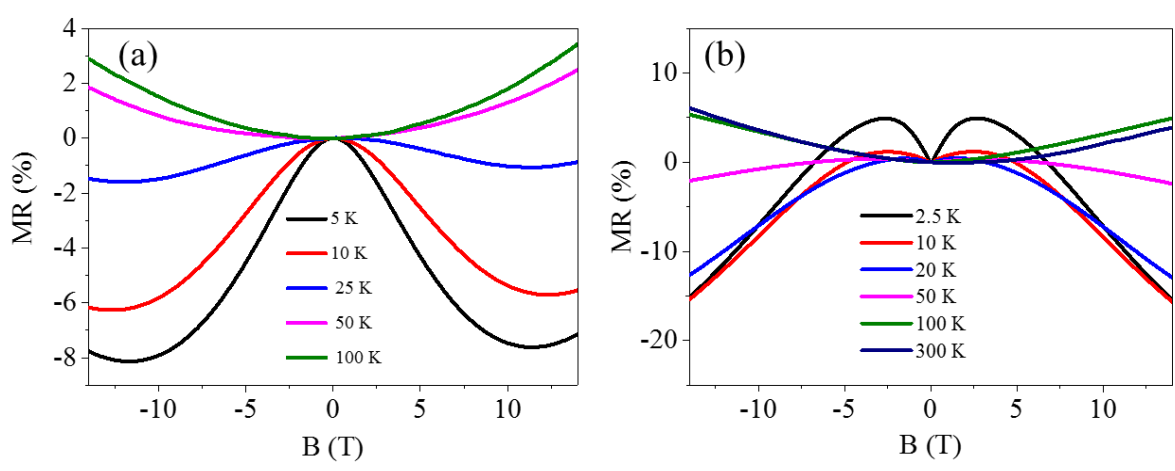

Figure. 4

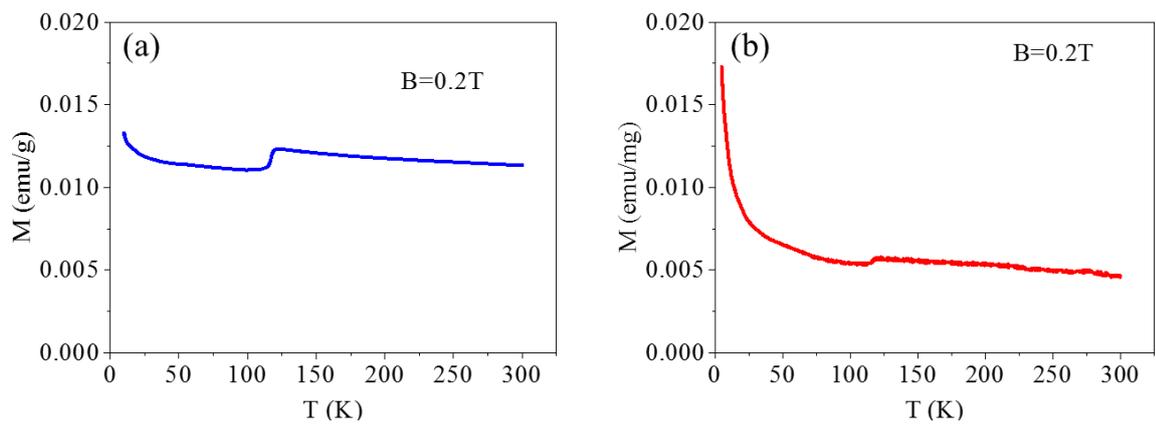

Figure 5.

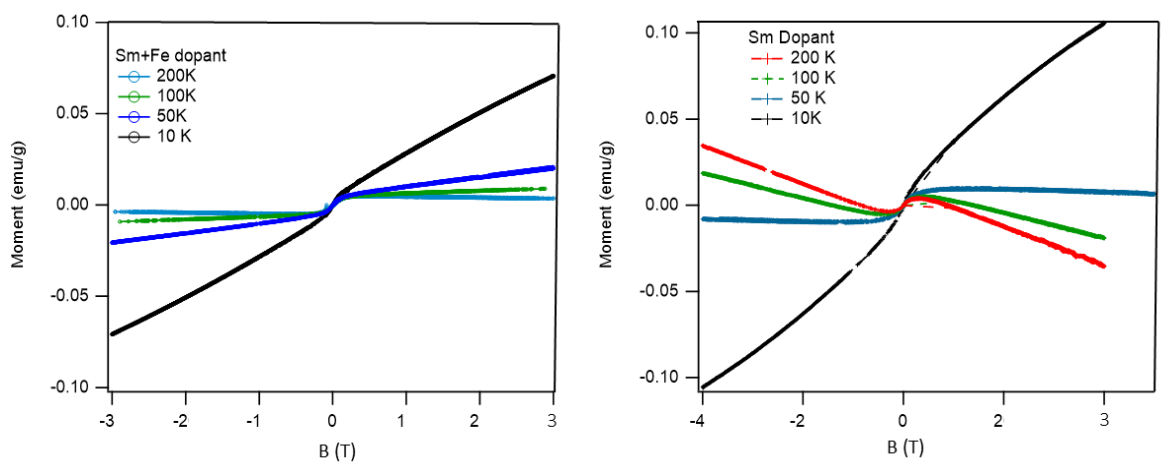